\documentclass[epjc3]{svjour3} 

\usepackage[margin=25mm]{geometry}
\usepackage{amsmath,amssymb} 

\def\hH{\hat{H}}%
\def\hn{\hat{n}}%
\def\hrho{\hat{\rho}}%
\def\betaph{\beta_{\mathrm{ph}}}%
\def\Tph{T_{\mathrm{ph}}}%
\newcommand{\Tr}[1]{\mathrm{Tr}\left[#1\right]}
\def\Eref{E_{\mathrm{ref}}}%
\def\Bzeta{\zeta_{\mathrm{B}}}
\def\Hzeta{\zeta_{\mathrm{H}}}

\begin{document}

\title{Thermodynamics of the independent harmonic oscillators with different frequencies in the Tsallis statistics}
\author{Masamichi Ishihara\thanksref{e1,addr1}}
\thankstext{e1}{email: m\_isihar@koriyama-kgc.ac.jp}
\institute{Department of Domestic Science, Koriyama Women's University, Koriyama, Fukushima, 963-8503, JAPAN \label{addr1}}


\abstractdc{
  We study the thermodynamic quantities in the system of the $N$ independent harmonic oscillators
  with different frequencies in the Tsallis statistics of the entropic parameter $q$ ($1<q<2$) with escort average.
  The self-consistent equation is derived, and the physical quantities are calculated with the physical temperature.
  It is found that the number of oscillators is restricted below $1/(q-1)$.
  The energy, the R\'enyi entropy, and the Tsallis entropy are obtained by solving the self-consistent equation approximately
  at high physical temperature and/or for small deviation $q-1$.
  The energy is $q$-independent at high physical temperature when the physical temperature is adopted,
  and the energy is proportional to the number of oscillators and physical temperature at high physical temperature.
  The form of the R\'enyi entropy is similar to that of von-Neumann entropy,  
  and the Tsallis entropy is given through the R\'enyi entropy.
  The physical temperature dependence of the Tsallis entropy is different from that of R\'enyi entropy. 
  The Tsallis entropy is bounded from the above,  
  while the R\'enyi entropy increases with the physical temperature. 
  The ratio of the Tsallis entropy to the R\'enyi entropy is small at high physical temperature.
}

\maketitle

\section{Introduction}

The various statistics have been proposed to describe the phenomena which show power-like distributions.
An extension of the Boltzmann-Gibbs statistics is the Tsallis statistics,
and the statistics has been applied in various branches of science \cite{TsallisBook}.
The escort average is often adopted to calculate the physical quantities in the Tsallis statistics.
The Tsallis statistics has the entropic parameter $q$, and
the statistics approaches the Boltzmann-Gibbs statistic as $q$ approaches one.

The entropic parameter $q$ is often restricted.
The normalizability of the probability requires that $q$ is less than two \cite{Tsallis-BJP39-overview}.
The parameter $q$ is also restricted because physical quantities are restricted
\cite{Tsallis-BJP39-overview,Ishihara2016:IJMPE,Bhattacharyya:2016,Bhattacharyya}.
For example, the energy density should be finite and the number of particles should be positive, 
and these requirements show that the maximum value of $q$ is smaller than two. 
The limitation of $q$ was also derived by using the conjugate variables theorem \cite{Umpierrez2021}.

Simple systems have been adopted to study the effects of statistics.
The classical gas model was adopted,
and it was found that the energy is proportional to the number of particles and the physical temperature 
\cite{Kalyana:2000,Abe-PLA:2001,S.Abe:physicaA:2001,Aragao:2003,Ruthotto:2003,Toral:2003,Suyari:2006,Ishihara:phi4,Ishihara:free-field,Ishihara:Thermodyn-rel}
in the Tsallis statistics.
It is also found that the number of the particles are restricted \cite{Abe-PLA:2001}.
The thermodynamic quantities for a classical harmonic oscillator was also calculated in the Tsallis statistics with escort average. 
The partition function was calculated and the energy was obtained \cite{Tsallis1998}.

The calculations of the thermodynamic quantities for the harmonic oscillators are required in the Tsallis statistics.
A field is decomposed into harmonic oscillators with different frequencies to calculate physical quantities. 
The results for the harmonic oscillators with different frequencies in the Tsallis statistics will be helpful 
to calculate physical quantities in various systems.

In this paper, we study the thermodynamic quantities in the system of the $N$ independent harmonic oscillators
in the Tsallis statistics of the entropic parameter $q$.
The range of $q$ is set between one and two in this study.
The escort average is employed to obtain physical values.
In Sec.~\ref{sec:review-Tsallis}, we briefly review the Tsallis statistics. 
In Sec.~\ref{sec:N-oscillators},
we study the $N$ independent harmonic oscillators with different frequencies.
The self-consistent equation is derived, and the equation is solved approximately.
The expression of the energy is obtained with physical temperature. 
The expressions of Tsallis and R\'enyi entropies are also obtained. 
The last section is assigned for conclusion.

\section{Brief review of the Tsallis statistics}
\label{sec:review-Tsallis}
The Tsallis statistics \cite{TsallisBook} is based on the Tsallis entropy $S_q^{(T)}$ with the entropic parameter $q$.
The entropy $S_q^{(T)}$ is defined by
\begin{align}
  S_q^{(T)} = \frac{1-\Tr{\hrho^q}}{q-1} ,
\end{align}
where $\hrho$ is the density operator.
The density operator $\hrho$ is obtained by extremizing $S_q^{(T)}$ under the normalization condition $\Tr \hrho = 1$ and the energy constraint: 
\begin{align}
  U = \frac{\Tr{\hrho^q \hH}}{\Tr{\hrho^q}} ,
  \label{eq:constraint:U}
\end{align}
where $U$ is the energy. The right-hand side of Eq.~\eqref{eq:constraint:U} is the escort average of the Hamiltonian $\hH$.

The density operator $\hrho$ in the Tsallis statistics with the escort average is obtained:
\begin{subequations}
\begin{align}
  & \hrho = \frac{1}{Z} \left( 1 - (1-q) \frac{\beta}{c_q} (\hH - U) \right)^{\frac{1}{1-q}}, \\
  & Z = \Tr{\left( 1 - (1-q) \frac{\beta}{c_q} (\hH - U) \right)^{\frac{1}{1-q}}}, \\
  & c_q = \Tr{\hrho^q},
\end{align}
\end{subequations}
where $\beta$ is the inverse temperature.
The partition function $Z$ is related to $c_q$:
\begin{align}
c_q = Z^{1-q}. 
\label{c_q:Z}
\end{align}
The inverse physical temperature is given by
\begin{align}
\betaph = \beta / c_q .
\end{align}
The physical temperature $\Tph$ is given as $1/\betaph$.

The thermodynamic quantities are calculated with the above density operator
for the $N$ independent harmonic oscillators with different frequencies in the following section.

\section{The independent harmonic oscillators with different frequencies}
\label{sec:N-oscillators}

\subsection{Derivation of self-consistent equation}
We attempt to derive the self-consistent equation by calculating $c_q$ in two ways.
One way is the method by using the relation $c_q = Z^{1-q}$ and the other way is the method by calculating $c_q = \Tr{\hrho^q}$ directly.
We obtain the self-consistent equation by equating these results.

We treat the $N$ independent harmonic oscillators with different frequencies. 
The Hamiltonian $\hH$ is 
\begin{align}
  \hH = \sum_{j=1}^N \omega_j \left( \hn_j + \frac{1}{2} \right),
\end{align}
where $\hat{n}_j$ is the number operator with the subscript $j$. 
We treat the above Hamiltonian in the Tsallis statistics of $1<q<2$: $(2-q)/(q-1)$, $1/(q-1)$, and $q/(q-1)$ are positive.

We introduce a parameter $\Eref$ and calculate the partition function $Z$:
\begin{align}
  Z &=
  \sum_{n_1, \cdots, n_N = 0}^{\infty} \left\{
  1+(q-1) \betaph \Bigg( \frac{\hbar}{2}(\omega_1+\cdots+\omega_N) -U \Bigg)
  + (q-1) \betaph \Bigg( \hbar \omega_1 n_1 + \cdots + \hbar \omega_N n_N \Bigg)
  \right\}^{\frac{1}{1-q}} \nonumber \\
  & = \left( (q-1)\betaph \Eref \right)^{\frac{1}{1-q}}
  \sum_{n_1, \cdots, n_N = 0}^{\infty}
  (\lambda_N + a_1 n_1 + \cdots + a_N n_N)^{\frac{1}{1-q}} , \label{eqn:Z}
\end{align}
where
\begin{subequations}
\begin{align}
& \lambda_N = \frac{1+(q-1) \betaph \Bigg( \displaystyle\frac{1}{2} \left(\displaystyle\sum_{i=1}^N \hbar\omega_i \right) -U \Bigg)}{(q-1) \betaph \Eref} , \\
& a_j := \frac{\hbar\omega_j}{\Eref} . 
\end{align}
\end{subequations}
Equation~\eqref{eqn:Z} is represented with Barnes zeta function $\Bzeta(s, \alpha| \vec{\omega}_N)$ (See Eq.~\eqref{def:Barnes-zeta}):
\begin{align}
  Z & = \left( (q-1)\betaph \Eref \right)^{\frac{1}{1-q}} \Bzeta\left(1/(q-1), \lambda_N | \vec{a}_N \right) \qquad \vec{a}_N = (a_1, a_2, \cdots, a_N) .
  \label{eqn:Z:Banes}
\end{align}
The condition $s>N$ for the parameters of the Barnes zeta function in the present case is 
\begin{align}
\frac{1}{q-1} > N. 
\end{align}
This means that the number of the oscillators is restricted.
We also calculate $c_q$ directly as
\begin{align}
  c_q = \Tr{\hrho^q}
  = Z^{-q} \left( (q-1)\betaph \Eref \right)^{\frac{q}{1-q}} \Bzeta\left(q/(q-1), \lambda_N | \vec{a}_N \right) .
  \label{cq-from-def}
\end{align}
From Eqs.~\eqref{c_q:Z}, \eqref{eqn:Z:Banes}, and \eqref{cq-from-def}, we have the following self-consistent equation:
\begin{align}
  \left( (q-1)\betaph \Eref \right) \Bzeta\left(1/(q-1), \lambda_N | \vec{a}_N \right)
  = \Bzeta\left(q/(q-1), \lambda_N | \vec{a}_N \right) .
  \label{self-consistent-eq}
\end{align}

We attempt to obtain the physical quantities by solving the self-consistent equation in the next subsection.

\subsection{Energy and entropies}
We attempt to find the expressions of physical quantities in this subsection. 
For $\lambda_N \gg 1$, we have the following expressions by using Eq.~\eqref{eqn:Bzeta:approximation}.
\begin{subequations}
\begin{align}
  &\Bzeta\left(1/(q-1), \lambda_N | \vec{a}_N \right)
  \sim
  \frac{(q-1)^N}
       {\left( \displaystyle\prod_{j=0}^{N-1} ((2-q)-j(q-1)) \right) 
         \left( \displaystyle\prod_{j=1}^{N} a_j \right)
         (\lambda_N)^{\frac{1}{(q-1)}-N}
       } , 
       \label{eqn:approximate1}\\
  &\Bzeta\left(q/(q-1), \lambda_N | \vec{a}_N \right)
  \sim  
  \frac{(q-1)^N}
       {\left( \displaystyle\prod_{j=0}^{N-1} (1-j(q-1)) \right) 
         \left( \displaystyle\prod_{j=1}^{N} a_j \right)
         (\lambda_N)^{\frac{q}{(q-1)}-N}
       } . 
       \label{eqn:approximate2}
\end{align}
\end{subequations}
In \ref{Barnes-zeta}, the above approximated expression for the Barnes zeta function is given by
using the approximated expression for the Hurwitz zeta function given in \ref{hurwitz-zeta}.
We use these expressions of $\Bzeta$ to solve Eq.~\eqref{self-consistent-eq} approximately.

\subsubsection{Expression of the energy}
We attempt to calculate the energy $U$ by solving Eq.~\eqref{self-consistent-eq}.
Substituting Eqs.\eqref{eqn:approximate1} and \eqref{eqn:approximate2} into Eq.~\eqref{self-consistent-eq},
we have
\begin{align}
U = 
\frac{\Tph}{(q-1)}
\left( 1 - \frac{\displaystyle\prod_{j=0}^{N-1} ((2-q)-j(q-1)) }{ \displaystyle\prod_{j=0}^{N-1} (1-j(q-1)) } \right)
+
\sum_{i=1}^N \frac{\hbar\omega_i}{2} 
,
\qquad N < \frac{1}{(q-1)}
. 
\label{expression-U}
\end{align}
It is noted that Eq.~\eqref{expression-U} does not contain $\Eref$. 
We obtain easily 
\begin{align}
  \frac{\displaystyle\prod_{j=0}^{N-1} ((2-q)-j(q-1)) }{ \displaystyle\prod_{j=0}^{N-1} (1-j(q-1)) } = 1 - N(q-1) . 
\end{align}
By substituting the above expression, we have the following expression of $U$:
\begin{align}
  U = N \Tph + \frac{1}{2} \sum_{i=1}^N \hbar\omega_i, \qquad N < \frac{1}{(q-1)} . 
  \label{eqn:U}
\end{align}
Equation \eqref{eqn:U} is the well-known form of the energy $U$ in the Boltzmann-Gibbs statistics.
It is possible to evaluate $\lambda_N$ by using Eq.~\eqref{eqn:U}:
\begin{align}
\lambda_N = \frac{1-N (q-1)}{(q-1)\betaph \Eref} .
\label{eqn:lambdaN:eval}
\end{align}
The numerator of the right-hand side of Eq.~\eqref{eqn:lambdaN:eval} is positive,
because $N (q-1)$ is less than one.
Therefore the condition $\lambda_N \gg 1$ is satisfied for $(q-1) \betaph \Eref \ll 1$:
the condition is satisfied at high physical temperature $\Tph$ and/or for small deviation $(q-1)$.

\subsubsection{Expressions of the entropies}

The Tsallis entropy $S_q^{(T)}$ is represented as
\begin{align}
S_q^{(T)} = \frac{1-c_q}{q-1} = \frac{1-Z^{1-q}}{q-1} .
\end{align}
The R\'enyi entropy $S_q^{(R)}$ is related to the Tsallis entropy:
\begin{align}
S_q^{(R)} = \frac{1}{1-q} \ln (1+(1-q)S_q^{(T)}) .
\end{align}
This equation is represented with $c_q$ as 
\begin{align}
S_q^{(R)} = \frac{1}{1-q} \ln c_q = \frac{1}{1-q} \ln e^{(1-q) \ln Z} = \ln Z .
\label{eqn:RenyiEntropy}
\end{align}

We calculate $Z$ approximately by using Eq.~\eqref{eqn:approximate1}.
\begin{align}
  Z =\frac{1}{
    \left( \displaystyle\prod_{j=0}^{N-1} ((2-q)-j(q-1)) \right)
    \left( \displaystyle\prod_{j=1}^{N} (\betaph \hbar \omega_j) \right)
    \left( 1 + (q-1) \betaph \left( \displaystyle\frac{1}{2} \displaystyle\sum_{i=1}^N (\hbar \omega_i) -U \right)\right)^{\frac{1}{q-1}-N}
  } .
  \label{eqn:Z:approximation}
\end{align}
Substituting Eq.~\eqref{expression-U} into Eq.~\eqref{eqn:Z:approximation}, we obtain
\begin{align}
  Z =  \frac{\Bigg( \displaystyle\prod_{j=0}^{N-1} (1-j(q-1)) \Bigg)^{\frac{1}{1-q}-N}}
       {\Bigg( \displaystyle\prod_{j=1}^{N} (\betaph \hbar \omega_j) \Bigg) \Bigg( \displaystyle\prod_{j=0}^{N-1} ((2-q)-j(q-1)) \Bigg)^{\frac{q}{1-q}-N}} . 
       \label{eqn:Z:approximation:q-1:rep}
\end{align}

We find the relation between $dU$ and $dS_q^{(R)}$. 
The R\'enyi entropy is given by $\ln Z$. For the fixed $N$ and $q$, we have
\begin{align}
  dS_q^{(R)} = d\ln Z = N \frac{d\Tph}{\Tph}. 
  \label{eqn:dSR}
\end{align}
With Eqs.~\eqref{eqn:U} and \eqref{eqn:dSR}, we have
\begin{align}
  dU = N d\Tph = \Tph dS_q^{(R)}.
\end{align}

The $q$-dependence of $S_q^{(R)}$ for small $q-1$ is obtained by 
expanding the logarithm of Eq.~\eqref{eqn:Z:approximation:q-1:rep} with respect to $q-1$.
We have
\begin{align}
  S_q^{(R)} = \ln Z = L_N(\Tph) + N + \frac{1}{2} N (q-1) + O((q-1)^2), 
  \label{eqn:lnZ} 
\end{align}
where $L_N(\Tph)$ is defined by 
\begin{align}
L_N(\Tph) := \sum_{j=1}^{N} \ln \Big(\frac{\Tph}{\hbar \omega_j}\Big) .
\end{align}
The same equation can be obtained by substituting Eq.~\eqref{eqn:U} into Eq.~\eqref{eqn:Z:approximation}.
We remember that $N(q-1)$ is less than one.

We obtain the ratio of $S_q^{(T)}$ to $S_q^{(R)}$.
Hereafter we omit the argument $\Tph$ of $L_n$ for simplicity.
For $L_N \gg N$, we have
\begin{subequations}
  \begin{align}
    &S_q^{(R)} \sim L_N,   \label{eqn:ST-highTph} \\
    &S_q^{(T)} \sim \frac{1-e^{-(q-1) L_N}}{q-1}. \label{eqn:SR-highTph}
  \end{align}
\end{subequations}
The ratio $S_q^{(T)}/S_q^{(R)}$ is
\begin{align}
\frac{S_q^{(T)}}{S_q^{(R)}} \sim \frac{1-e^{-(q-1) L_N}}{(q-1)L_N}.
\end{align}
The ratio $S_q^{(T)}/S_q^{(R)}$ is approximately $1/((q-1) L_N)$ at sufficiently high physical temperature which satisfies $(q-1) L_N \gg 1$.
This ratio is $1 - (q-1) L_N/ 2$ for $(q-1) L_N \ll 1$, 
though the condition $L_N \gg N$ is required to obtain the expressions, Eqs.~\eqref{eqn:ST-highTph} and \eqref{eqn:SR-highTph}.

\section{Conclusions}
\label{sec:conclusion}
We studied the thermodynamic quantities in the system of the $N$ independent harmonic oscillators
with different frequencies $\omega_j$ in the Tsallis statistics of the entropic parameter $q$ ($1<q<2$).
The number of the oscillators $N$ was fixed and the escort average was adopted in this study.
We derived the self-consistent equation,
and the expressions of physical quantities with the physical temperature were obtained.
We obtained the partition function $Z$, the energy $U$, the R\'enyi entropy $S_q^{(R)}$, and the Tsallis entropy $S_q^{(T)}$
by solving the self-consistent equation approximately at high physical temperature  $\Tph$ and/or for small deviation $q-1$.

It was found from the condition for the parameters of the Barnes zeta function 
that the number of harmonics oscillators $N$ is less than $1/(q-1)$.
The restriction of the number of the harmonic oscillators exists, 
as the restriction was previously given for the classical gas \cite{Abe-PLA:2001}. 
As expected, the supremum $1/(q-1)$ goes to infinity when $q$ approaches one.

The energy $U$ is $q$-independent at high physical temperature when the physical temperature is adopted. 
The energy is proportional to the number of harmonic oscillators $N$ and the physical temperature $\Tph$ at high physical temperature
when the vacuum term is ignored: 
the expression of the energy is the well known expression, ${U = N \Tph + \sum_j \hbar\omega_j/2}$. 
The R\'enyi entropy $S_q^{(R)}$ is the sum of the values for the independent harmonic oscillators at high physical temperature.
The R\'enyi entropy with the same frequency, $\omega \equiv \omega_1 = \cdots = \omega_N$, is given by $N \ln(\Tph/(\hbar \omega))$
which is well-known expression for the $N$ independent harmonic oscillators with the same frequency. 
The Tsallis entropy $S_q^{(T)}$ was obtained through the R\'enyi entropy.  
The variation for the R\'enyi entropy is simply given as $dS_q^{(R)} = N d\Tph$ for the fixed $N$,
and the well-known relation between $dU$ and $dS_q^{(R)}$ is also obtained: $dU = \Tph dS_q^{(R)}$. 

The physical temperature dependence of the Tsallis entropy is different from that of the R\'enyi entropy. 
The R\'enyi entropy contains the term that is $\sum_{j=1}^N \ln(\Tph/(\hbar \omega_j))$.  
Therefore, the R\'enyi entropy increases with the physical temperature, and is unbounded from the above. 
In contrast, the Tsallis entropy increases with the physical temperature, and is bounded from the above.  
The ratio of the Tsallis entropy to the R\'enyi entropy, $S_q^{(T)}/S_q^{(R)}$, is small at high physical temperature. 
The difference between the Tsallis and R\'enyi entropies is large at high physical temperature.

The system of the independent harmonic oscillators with different frequencies is basic,
and the results in this study will give the insight on other physical systems.
The author believes that
the present study will be helpful for the reader to study the system represented with oscillators
in unconventional statistics such as the Tsallis statistics.

\appendix
\section{Approximate expression of Hurwitz zeta function}
\label{hurwitz-zeta}
The Hurwitz zeta function \cite{Espinosa,Bordag,Shpot} is defined by
\begin{align}
  & \Hzeta(s,\alpha) := \sum_{n=0}^{\infty} \frac{1}{(\alpha + n)^s} . 
  \label{def:Hurwitz-zeta}
\end{align}
We treat the case of $s>1$ and $\alpha > 0$ in this appendix.

Let $B_n(x)$ be Bernoulli polynomials which are defined by 
\begin{align}
\frac{te^{xt}}{e^t-1} = \sum_{n=0}^{\infty} B_n(x) \frac{t^n}{n!} .
\end{align}
The Bernoulli number $B_n$ in this paper is defined
\footnote{
  The Bernoulli number $B_n$ is often defined as $B_n(x=0)$.
  It may worth to mention that $B_n(x=0) = B_n(x=1)$ for $n \neq 1$. 
}
by
\begin{align}
B_n := B_n(x=1).
\end{align}

We use the Euler-Maclaurin formula.
Let $a$ and $b$ be integer with $a < b$ and let $f(x)$ be continuously differentiable for $M$-times.
The Euler-Maclaurin formula is 
\begin{align}
  \sum_{n=a}^b f(n)
  = & \int_a^b dx f(x) + \frac{1}{2} (f(b)+f(a))
  + \sum_{k=1}^{M-1} \frac{B_{k+1}}{(k+1)!} (f^{(k)}(b) - f^{(k)} (a)) \nonumber \\
  & - \frac{(-1)^M}{M!} \int_a^b dx B_M(x-[x]) f^{(M)}(x) , 
\end{align}
where $f^{(k)}$ is $k$-th derivative and $[x]$ is the Gauss symbol (the floor function).

We attempt to find the expression of $\Hzeta(1+z, \alpha)$ for $z>0$
by using the Euler-Maclaurin formula. 
The right-hand side of Eq.~\eqref{def:Hurwitz-zeta} is an infinite series.
Therefore, we first consider the following finite series:
\begin{align}
  \zeta_{H,m}(s,\alpha) = \sum_{n=0}^m \frac{1}{(\alpha + n)^s} .
\end{align}
We set $f(x)$ as $1/(\alpha + x)^{1+z}$ and apply the Euler-Maclaurin formula.
By taking the limit $m \rightarrow \infty$, we have the expression of $\Hzeta(1+z,\alpha)$.
The integral part converges when $\alpha$ is positive.
We finally obtain
\begin{align}
  \Hzeta(1+z, \alpha)
  &= \frac{1}{z\alpha^z} + \frac{1}{2\alpha^{1+z}} + \sum_{k=1}^{M-1} \frac{(-1)^{k+1} B_{k+1}}{(k+1)!} \frac{\Gamma(z+k+1)}{\Gamma(z+1)} \frac{1}{\alpha^{z+k+1}}
  \nonumber \\
  & \quad - \frac{(-1)^M}{M!} \int_0^{\infty} dx B_M(x-[x]) f^{(M)}(x) \qquad (z>0, \alpha > 0) .
\label{eqn:Hzeta}
\end{align}
The function $\Hzeta(1+z, \alpha)$ can be rewritten \cite{Boumali2014}.
For example, $\Hzeta(1+z, \alpha)$ is given by 
\begin{align}
  \Hzeta(1+z, \alpha)
  &= \frac{1}{z\alpha^z} + \frac{1}{2\alpha^{1+z}} + \frac{1}{z} \sum_{k=2}^{M} \frac{B_k}{k!} \frac{\Gamma(z+k)}{\Gamma(z)} \frac{1}{\alpha^{z+k}}
  \nonumber \\
  & \quad - \frac{(-1)^M}{M!} \int_0^{\infty} dx B_M(x-[x]) f^{(M)}(x) \qquad (z>0, \alpha > 0), 
\end{align}
because $B_{2n+1}$ is zero for $n \ge 1$ and $\Gamma(z+1) = z \Gamma(z)$.

It is possible to estimate the integral of Eq.~\eqref{eqn:Hzeta} by setting $M$.
For example, the upper value of the integral with $M=2$ is estimated:
\begin{align}
  \left| \frac{1}{2!} \int_0^{\infty} B_2(x-[x]) f^{(2)}(x) \right| \le  \frac{C_2}{2!} \int_0^{\infty}  \left| f^{(2)}(x) \right|, 
\end{align}
where $C_2$ is the maximum value of $|B_2(x)|$ in the range of $0 \le x \le 1$.

From Eq.~\eqref{eqn:Hzeta}, we find that the $\Hzeta(1+z, \alpha)$ for $\alpha \gg 1$ behaves
\begin{align}
  \Hzeta(1+z,\alpha) \sim \frac{1}{z \alpha^{z}}.
  \label{eqn:Hzeta:large}
\end{align}

\section{Approximate expression of Barnes zeta function}
\label{Barnes-zeta}
The Barnes zeta function \cite{Ruijsenaars:2000,Kirsten:2010} is defined by
\begin{align}
  & \Bzeta(s,\alpha|\vec{\omega}_N) := \sum_{n_1,\cdots,n_N=0}^{\infty} \frac{1}{(\alpha + \omega_1 n_1 + \cdots + \omega_N n_N)^s}
  \qquad \vec{\omega}_N = (\omega_1, \omega_2, \cdots, \omega_N), 
  \label{def:Barnes-zeta}
\end{align}
where $s>N$, $\alpha > 0$, and $\omega_j > 0$. 

We define $\Omega_N$ as
\begin{align}
\Omega_N := \alpha + \omega_1 n_1 + \cdots + \omega_N n_N .
\end{align}
The function $\Bzeta$ is represented as 
\begin{align}
  \Bzeta(s,\alpha|\vec{\omega}_N)
  &= \sum_{n_1,\cdots,n_N=0}^{\infty} \frac{1}{(\Omega_N)^s}
  = \frac{1}{(\omega_N)^s} \sum_{n_1,\cdots,n_{N-1}=0}^{\infty} \sum_{n_N=0}^{\infty} \frac{1}{((\Omega_{N-1}/\omega_N)+n_N)^s}
  \nonumber \\
  &= \frac{1}{(\omega_N)^s} \sum_{n_1,\cdots,n_{N-1}=0}^{\infty} \Hzeta(s, \Omega_{N-1}/\omega_N) .
\end{align}
We have $\Hzeta(1+z,\alpha) \sim 1/(z \alpha^z)$ for $\alpha \gg 1$.
Therefore, for sufficiently large $\alpha$, we have
\begin{align}
  \Bzeta(1+z,\alpha|\vec{\omega}_N) \sim  \frac{1}{(\omega_N)^{1+z}} \sum_{n_1,\cdots,n_{N-1}=0}^{\infty} \frac{1}{z (\Omega_{N-1}/\omega_N)^{z}}
  = \frac{1}{z \omega_N} \Bzeta(z, \alpha|\vec{\omega}_{N-1}) .
\label{eq:recurrence-relation}
\end{align}
By using the recurrence relation, Eq.~\eqref{eq:recurrence-relation}, we have the approximate expression of $\Bzeta$ for $\alpha \gg 1$:
\begin{align}
  \Bzeta(1+z, \alpha|\vec{\omega}_N)
  &\sim  \frac{1}{z (z-1) \cdots (z-(N-1))} \frac{1}{\omega_1 \omega_2 \cdots \omega_N} \frac{1}{\alpha^{z-(N-1)}}
  \nonumber \\
  &= \frac{1}{\Bigg(\displaystyle\prod_{j=0}^{N-1} (z-j)\Bigg) \Bigg(\displaystyle\prod_{j=1}^{N} \omega_j \Bigg) \alpha^{z-(N-1)}} 
  \qquad\qquad (z - (N-1) > 0). 
  \label{eqn:Bzeta:approximation}
\end{align}
The condition $z - (N-1) > 0$ is rewritten as $1+z - N >0$.
This condition is equivalent to the condition $s>N$ with $s=1+z$ for $\Bzeta(s, \alpha|\vec{\omega}_N)$.


\end{document}